\documentclass[mathleft,fleqn,%
final,%
]{an}
\usepackage{amsmath}
\usepackage{amssymb,bm}
\usepackage{graphicx}
\usepackage[varg]{txfonts}
\usepackage{natbib}
\bibpunct{(}{)}{;}{a}{}{,}
 \usepackage{flushend} 
\newcommand{\Bep}{B_\mathrm{ep}}
\newcommand{\Bet}{B_\mathrm{et}}

\begin{document}

\Pagespan{1082}{1089}
\Yearpublication{2016}%
\Yearsubmission{2016}%
\Month{12}%
\Volume{337}%
\Issue{10}%
\DOI{10.1002/asna.201612446}%

\title{Helical kink instability in the confined solar eruption on 2002 May 27}

\author{A.~Hassanin\inst{1,2}\fnmsep\thanks{Corresponding author: {hassanin@uni-potsdam.de}} 
   \and B.~Kliem\inst{1} 
   \and N.~Seehafer\inst{1}}

\titlerunning{Confined solar eruption on 2002 May 27}
\authorrunning{A. Hassanin, B. Kliem, \& N. Seehafer}
\institute{Institute of Physics and Astronomy, University of Potsdam, 14476 Potsdam, Germany
      \and Department of Astronomy, Space Science \& Meteorology, Faculty of Science, University of Cairo, Cairo, Egypt}

\received{2016 Sep 27}
\accepted{2016 Sep 30}
\publonline{.....}

\keywords{Instabilities -- magnetohydrodynamics (MHD) -- Sun: corona -- Sun: coronal mass ejections (CMEs) -- Sun: flares}

\abstract{This paper presents an improved MHD modeling of the confined filament eruption in solar active region NOAA~9957 on 2002 May~27 by extending the parametric studies of the event in \citet{Torok&Kliem2005} and \citet{Hassanin&Kliem2016}. Here the initial flux rope equilibrium is chosen to possess a small apex height identical to the observed initial filament height, which implies a more realistic inclusion of the photospheric line tying. The model matches the observations as closely as in the preceding studies, with the closest agreement again being obtained for an initial average flux rope twist of about $4\pi$. Thus, the model for strongly writhing confined solar eruptions, which assumes that a kink-unstable flux rope in the stability domain of the torus instability exists at the onset of the eruption's main acceleration phase, is further substantiated.}

\maketitle

\section{Introduction}

Confined solar eruptions consist of a filament or prominence eruption and usually an associated flare, but do not evolve
into a coronal mass ejection (CME); rather, the moving plasma is halted in the solar corona and usually seen to fall back.
Flares not associated with any rising material also belong to this category. Clarifying the factors
that determine the confined vs.\ ejective nature of an eruption is necessary to establish a comprehensive model of solar
eruptions. This is a key task in the study of space weather.

The first detailed observations of a confined filament eruption were obtained on 2002 May~27 by the \textsl{TRACE} satellite 
\citep{Handy&al1999} in the 195~{\AA} band (\citealt{HJi&al2003}; \citealt{Alexander&al2006}), see Fig.~\ref{f:shape}. 
The filament developed a strongly helical shape during its rise to the terminal height of ${\sim}\,0.1 R_{\odot}$, where its top part became
considerably distorted. Brightenings initially occurred at the top side of the rising filament, and its threads subsequently
reconnected with the overlying field, which led to the filament's complete disintegration. Bright footpoint sources under the
filament at the peak of the hard X-ray flux suggest that magnetic reconnection under the rising filament, similar to that 
in a CME, also occurred. Flare loops appeared about one hour after the start of the event and showed two very unusual properties: a substantial initial height (${\sim}\,0.05R_\odot$) and indications of twist. 

These properties support a model for confined eruptions, which assumes that the erupting flux consists of
a twisted flux rope susceptible to the helical kink instability but stable with regard to the torus
instability (\citealt{Torok&Kliem2005}, hereafter TK05; \citealt{Hassanin&Kliem2016}, hereafter HK16). 
The helical kink mode produces the observed helical
shape and steepens a helical current sheet in the interface to the ambient flux, where reconnection commences.
The interaction with the overlying field halts the rise if the torus instability is prevented by a slow decrease
of the field strength with height. This also distorts the rising flux rope and facilitates the reconnection in the helical
current sheet, which cuts the flux rope completely. Finally, if the kink-unstable rope develops a sufficient writhe, its
reconnected halves are interlinked and begin to reconnect a second time in a vertical current sheet that forms between them.
This reconnection reforms a far less twisted flux rope low in the corona and arcade-like (but weakly twisted)
overlying flux in the upward reconnection outflow, which is observed as flare loops. Detailed quantitative comparisons
between the observations and a parametric simulation study of the model showed good agreement and constrained the relative
magnitudes of the external poloidal (strapping) and toroidal (shear) field components to a value around unity, $\Bep/\Bet\!\approx\!1$,
and the initial flux rope twist to $\Phi\!\approx\!4\pi$.

The numerical modeling by TK05 and HK16 realized only a relatively poor matching of the initial filament height, 
which was too high in the model by a factor of ${\approx}\,1.3$.
Although this mismatch appears rather modest by number, the shape of the relatively high arching model flux rope implies a weaker
effect of the photospheric line tying compared to reality. To see this, one can represent the flux rope by a section of a torus
and consider a flux surface at a certain radial distance $a_1$ from the toroidal axis. The flatter the rope, the larger the
two end sections of the rope axis that are not fully enclosed by the flux surface. The field lines at radius $a_1$ in these
sections are arch-shaped, i.e., they do not pass under the rope axis, different from the middle section of the axis.
This arch-shaped flux realizes the stabilizing line tying effect, whereas the flux in the middle section contributes to the
destabilizing hoop force. Therefore, we here complement the parametric study of HK16 to test if the model can
still yield a good match with the observed data if the line tying is included much more realistically. We submerge the
center point of the toroidal flux rope deeper below the photosphere to obtain a flatter coronal rope section. All basic assumptions
and the other parameter settings are kept the same as in TK05 and HK16.

\section{Numerical model}

The plasma beta in active regions at low heights in the corona, where solar eruptions originate, is very small, so that in equilibrium the magnetic field is nearly force-free. Furthermore, the component of the pressure gradient perpendicular to the magnetic field  is generally much smaller than the Lorentz force if the equilibrium is lost and an eruption develops. The same is true for the gravitational force component perpendicular to the field. In the main acceleration phase of solar eruptions, any plasma motions along the field are secondary to the expulsion of the plasma.
Therefore, a reasonable approximation in the modeling of the main acceleration phase consists in starting from a force-free equilibrium and neglecting the thermal pressure and gravity in the dynamical evolution. Since we do not intend to model the draining of the filament plasma along the field  after the upward motion of the filament has stopped, the zero-beta approximation of the compressible ideal MHD equations in the absence of gravity is used,
\begin{eqnarray}
\partial_t\rho &=&
-\bm{\nabla\cdot}(\rho\,\bm{u})\,,      \label{eq_rho}\\
\rho\,\partial_{t}\bm{u}&=&
-\rho\,(\,\bm{u\cdot\nabla})\,\bm{u}
+\bm{J\times B} 
+\bm{\nabla\cdot\mathsf{T}}\,,                     \label{eq_mot}\\
\partial_{t}\bm{B}&=&
\bm{\nabla \times} (\bm{u\times B})\,, ~\mathrm{and}          \label{eq_ind}\\
\bm{J}&=&\mu_0^{-1}\,\bm{\nabla\times B}\,.              \label{eq_cur}
\end{eqnarray}
Here, $\bm{\mathsf{T}}$ is the viscous stress tensor 
($\mathsf{T}_{ij}=\rho\,\nu\,[\partial u_{i}/\partial x_j+
\partial u_{j}/\partial x_i-
(2/3)\delta_{ij}\,\bm{\nabla\cdot u}]$,
with $\nu$ denoting the kinematic viscosity).

The filament is modeled as a line tied force-free flux rope equilibrium according to \citeauthor{Titov&Demoulin1999} (\citeyear{Titov&Demoulin1999}, hereafter TD99). This initial configuration is composed of three major components. The first component is the upper section of a toroidal current channel with major and minor radii $R$ and $a$, respectively, placed in the $y$-$z$ plane such that its center is submerged below the photosphere, $\{z=0\}$, by a depth $d$. The channel carries a total ring current $I$. An external poloidal field $\Bep$ forms the second component. It is introduced by a pair of magnetic sources of strength $\pm q$ below the photosphere, placed at the symmetry axis of the torus at a distance $\pm L$ from the torus plane. The corresponding flux concentrations in the magnetogram at the positions $x=\pm L$ resemble a pair of sunspots in a bipolar active region. Finally, the third component is an external toroidal field $\Bet$, created by a line current $I_{o}$ which runs below the photosphere along the symmetry axis of the torus. In this model, all ambient flux above the photosphere is current-free. The initial density distribution is chosen as $\rho_{o}(x)=|B_{0}(x)|^{\frac{3}{2}}$, as in our previous work. This yields a height profile of the Alfv\'en velocity consistent with radio observations \citep{Vrsnak&al2002} and previous modeling \citep{Regnier&al2008}. The plasma is at rest initially, except for a small upward velocity perturbation applied only at the flux rope apex for a few Alfv\'en times (see below and HK16 for the details). This perturbation generates an upward kink of the flux rope apex and launches the helical kink instability if the initial configuration is unstable. 

We use a static but stretched grid, which allows for high resolution of the flux rope with relatively remote boundaries. The velocity is kept at zero at the boundaries. This represents the dense photosphere at the bottom boundary, $\{z=0\}$. As a result, the vertical component of the magnetogram does not change there, and the flux rope is line tied to the photosphere. This setting for the velocity also implements closed side and top boundaries. The initial apex height of the flux rope, the field strength, density, and Alfv\'en velocity at its magnetic axis, and the resulting Alfv\'en time are used to normalize the variables, and
Eqs.~(\ref{eq_rho})--(\ref{eq_cur}) are integrated using a modified Lax-Wendroff scheme (cf. TK05 and HK16).

\section{Simulation of the confined eruption}\label{s:simulations}

\begin{figure*}[!t]                                                
	\centering
	\includegraphics[height=0.868\textheight]{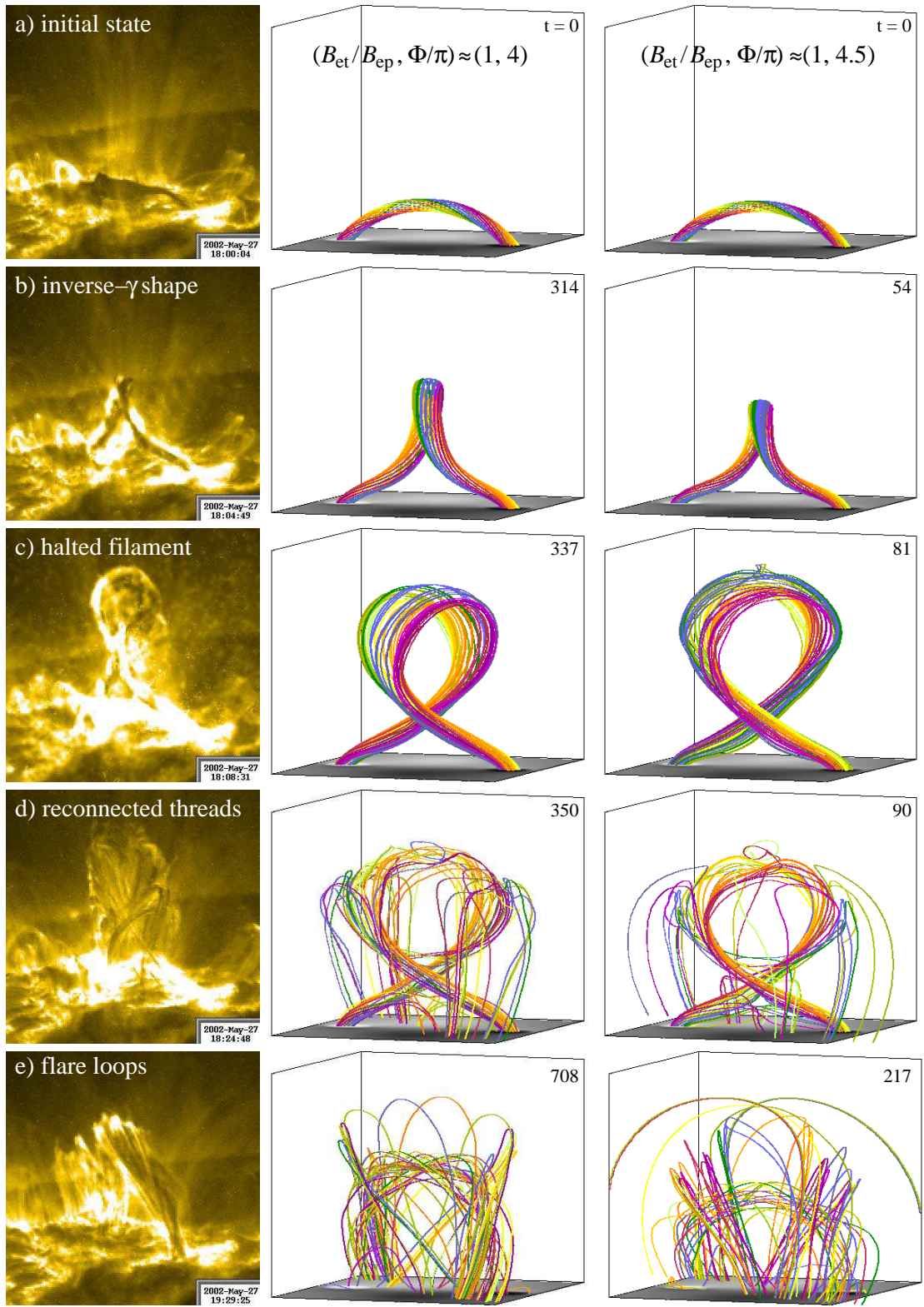} 
	\caption[caption]{Comparison of the main features of the confined filament eruption in active region NOAA~9957 at about 18~UT on 2002 May~27 observed by \textsl{TRACE} at 195~{\AA} (\textit{left column}) with Cases~1 (\textit{middle column}) and 2 (\textit{right column}). Only the central part of the box, the volume $6^3$, is shown (lengths are normalized by the initial apex height $h_0$ of the flux rope). The magnetogram, $B_z(x,y,0,t)$, is included in gray scale. Panels (a)--(d) show magnetic field lines outlining the core of the kink-unstable flux rope (with starting points on a circle of radius $r=a/3$, centered at the rope axis). Panel (e) shows ambient field lines after the two main reconnection phases.}
	\label{f:shape}
\end{figure*}

For a torus of major radius $R$ with the center point submerged by depth $d$, the apex height of the toroidal axis is $h_0=R-d$ and the footpoint distance of the coronal section is $D_\mathrm{f}=2(R^2-d^2)^{1/2}$. The projected initial and terminal apex heights of the filament in the considered event were determined by \citet{HJi&al2003} from the \textsl{TRACE}~195~{\AA} images to be $h_0=17.4$~Mm and $h_\infty=84.4$~Mm, respectively, giving $h_\infty/h_0=4.85$. These values are close to the true values because the event occurred close to the (west) limb and the rise did not indicate any strong non-radial direction. HK16 estimated $h_\infty/D_\mathrm{f}\approx1.1$ from the \textsl{TRACE} images.
Hence, $h_0/D_\mathrm{f}\approx1.1/4.85=0.23$.

Let the quantity $\kappa$ be defined by $\kappa=(2h_0/D_\mathrm{f})^2$. Inserting the expressions for $h_0$ and $D_\mathrm{f}$ into this definition and resolving for $d/R$ gives $d/R=(1-\kappa)/(1+\kappa)$.
The above observational estimates fix  $\kappa$ to a value of $0.21$.
Thus, $d/R=0.66$. For the value of $R=110$~Mm assumed in HK16 prior to the scaling of the simulations to the observations, this yields $d=72$~Mm, while actually a depth  $d$ of only 50~Mm was used in that study.
Requiring further that $h_0$ matches the observationally estimated value, thus fixing both $d/R$ and $R\!-\!d$, we obtain $R=51$~Mm and $d=33.6$~Mm. 

In order to model the confined nature of the eruption, flux rope equilibria in the stability domain of the torus instability are considered. This requires that the external field, especially the external poloidal field, decreases sufficiently slowly with height at and above the initial flux rope position (\citealt{Roussev&al2003}; \citealt{Kliem&Torok2006}). The vertical scale length of $\Bep$ scales linearly with the ``sunspot distance'' $L$ in the model, so that a sufficiently large value of $L$ guarantees torus stability and confinement. For each $L$, the strength of $\Bep$ at the position  of the flux rope (at its magnetic axis) is determined by the equilibrium condition (see, e.g., TD99), which fixes $q$ for given $I$, or vice versa. Thus, if the geometry of the flux rope ($R$ and $a$) and the ring current $I$ are set, $q$ will be determined for each $L$ such that $\Bep$ at the magnetic axis attains the equilibrium value, which is given by $R$, $a$, and $I$ but does not depend on $L$. However, different $L$ yield a different decrease of $\Bep$ with height, i.e., a different amount of overlying flux. There is a critical value $L_\mathrm{cr}$ for the onset of the torus instability, with instability for $L<L_\mathrm{cr}$. For $\Bet=0$, one finds $L_\mathrm{cr}=R$ for the bipolar $\Bep$ of the TD99 equilibrium if the canonical value for the critical decay index of the external poloidal field for the onset of the torus instability is used, $n = -\mathrm{d}\,\log\,\Bep(R)/\mathrm{d}\,\log\,R = 3/2$. The critical value of $L$ decreases for increasing $\Bet>0$ (which acts stabilizing), but this dependence has not yet been quantified. Since the overlying flux decreases with decreasing $L$, the terminal height of an erupting flux flux rope increases with decreasing $L$. 

The strength of $\Bet$ in the simulations presented here is taken to be the optimum one found in our previous extended parametric study of the event, given by $I_0=3.15\times10^{12}$~A. This yields $\Bet/\Bep\approx1$ at the magnetic axis of the flux rope. 

In order to initiate an eruption and model the observed helical shape, the initial flux rope twist is set near the critical value for the onset of the helical kink instability. This value is $\Phi_\mathrm{cr}\approx2.5\pi$ for a line tied and uniformly twisted tube with $\Bet=0$ (\citealt{Hood&Priest1981}; \citealt{Einaudi&vanHoven1983}), but rises for increasing $\Bet>0$ and increasing non-uniformity of the radial twist profile. Systematic investigations of these parametric dependencies, especially the dependence on $\Bet$, have not yet been done. Studies of line tied, kink-unstable, arched flux ropes with aspect ratios $R/a\sim2\mbox{--}10$ and a range of values for $\Bet$, both representative of erupting solar flux, suggest that the critical twist may typically be near the value $\Phi_\mathrm{cr}\approx3.5\pi$ (\citealt{Fan&Gibson2003}; \citealt{Torok&al2004}; TK05; \citealt{Kliem&al2012}; HK16). Therefore, we ran the simulations with the initial twist set to $\Phi=3.5\pi$, $4\pi$, and $4.5\pi$, which is accomplished by adjusting the minor radius $a$. These twist values are averaged over the cross section of the current channel. It turned out that, for the given geometry ($a/R$ and $d/R$) and the given $\Bet$, the twist $\Phi=3.5\pi$ lies well in the kink-stable domain of the parameter space, so that even a moderate upward initial velocity perturbation could not launch an eruption. Hence, we present and compare the simulation runs with $\Phi=4\pi$ and $\Phi=4.5\pi$. The corresponding values $L=30$~Mm and $L=37$~Mm, respectively, were determined by trial and error such that the observed terminal height of the event is closely matched. 

The range of initial twists considered here is consistent with the results of \citet{YGuo&al2013}, \citet{TLi&JZhang2015}, and \citet{Liu&al2016}, who found values up to about $4\pi$ in other eruptive solar events. The parametric simulation study of the present event in HK16 favored this same value but yielded a reasonable agreement with the observations also for $\Phi=4.5\pi$.

\section{Results}

\subsection{Shapes formed by instability and reconnection}

\begin{figure*}[!t]                                                
	\centering
	\includegraphics[height=0.7\textheight,width=0.525\textwidth]{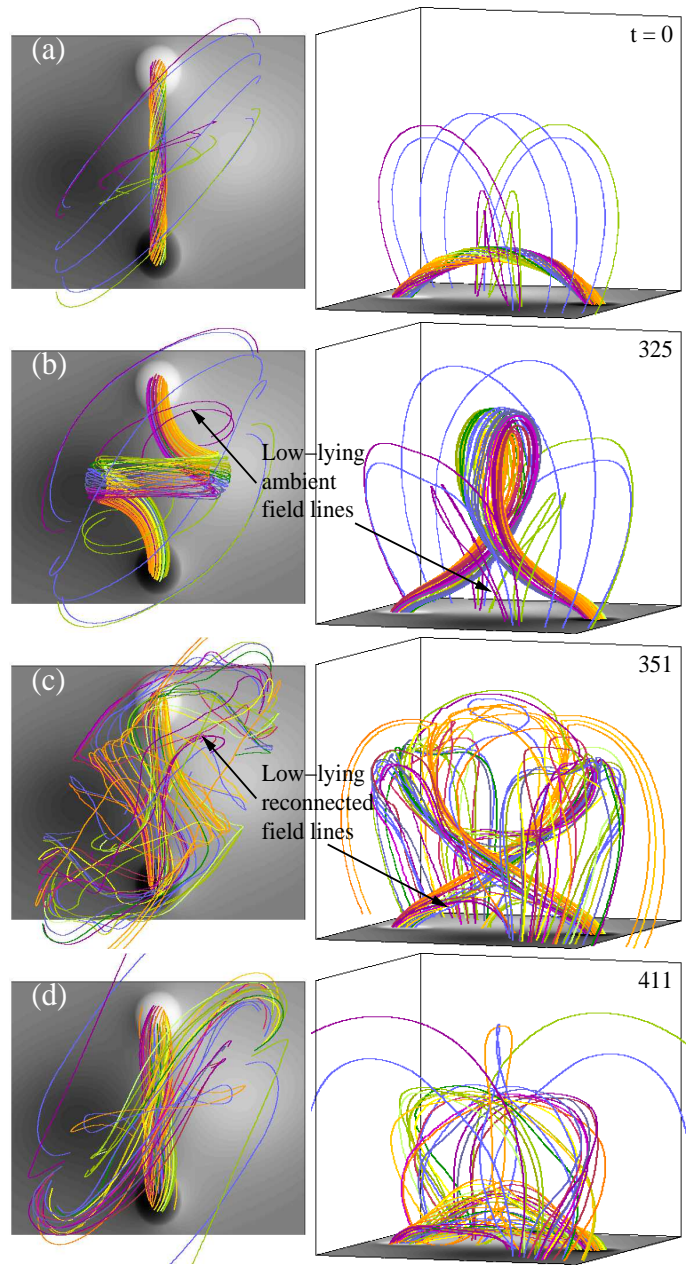} 
	\caption{Top and side views represent an overview of the reconnection in Case~1. The first main reconnection phase occurs in the helical current sheet, mainly between the top part of the erupted flux rope and the overlying flux (panels (c)). This eventually cuts the flux rope completely, producing two bundles of linked flux. Some additional reconnection with ambient flux occurs in the lower part of the helical current sheet (indicated in panels (c)). The second main reconnection phase occurs in a vertical current sheet that forms between the approaching, linked flux bundles, resulting in the reformation of a weakly twisted flux rope and an arch-shaped overlying flux (panels (d)). The latter is weakly twisted because it includes the part of the initial flux rope above the leg crossing point.}
	\label{f:reconnection}
\end{figure*} 

\begin{figure*}[!t]                                                
	\centering
	\includegraphics[width=\textwidth]{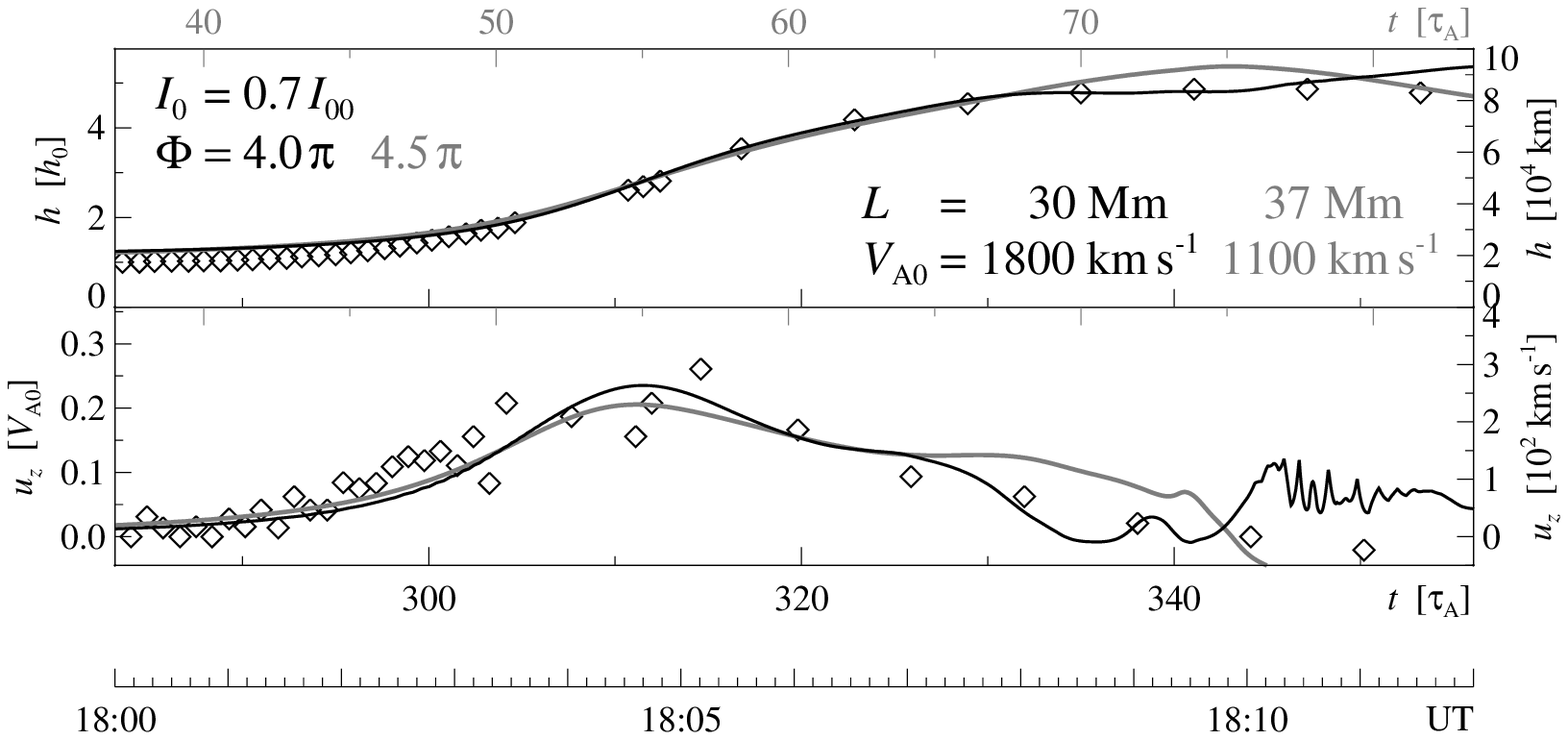} 
	\caption{Comparison of height and velocity of the fluid element initially at the apex of the flux rope's magnetic axis with the observed rise curve of the filament apex. \textit{Black (gray) lines and legends} refer to  Case~1 (Case~2). Smoothed height data from \citet{HJi&al2003} and smoothed, derived velocities are shown as diamonds. The simulation quantities are normalized by the initial apex height $h_0$, Alfv\'en velocity $V_\mathrm{A0}$ at the rope's magnetic axis, and the resulting Alfv\'en time $\tau_\mathrm{A}=h_0/V_\mathrm{A0}$. $I_{00}$ is the line current used in TK05.}
	\label{f:time_profile}
\end{figure*}

The two simulation runs analyzed in this paper are Case~1 with $\Phi=4\pi$ and $L=30$~Mm and Case~2 with $\Phi=4.5\pi$ and $L=37$~Mm. They are first compared with each other and with the \textsl{TRACE} observations based on the field line plots in Fig.~\ref{f:shape}, where panels (a)--(d) show flux near the axis of the flux rope ($r\le a/3$) and panel (e) shows ambient flux that has reconnected twice with the rope. The times for each case were selected such that the resulting shapes correspond to the four characteristic shapes that occurred in the event. 

The helical deformation (writhing) of the rising filament first produced an \textit{inverse-gamma shape} (Fig.~\ref{f:shape}(b)). Both cases match this shape quite well. The higher twist of Case~2 yields a slightly better match of the S-shape developed by the right (southern) filament leg. 

As the eruption approached and reached its terminal height, a strong \textit{distortion of the halted filament} occurred (Fig.~\ref{f:shape}(c)). The threads in the upper part of the filament were dragged to the side, which marked the beginning disintegration of the filament. The distortion results from the sideways oriented outflows of the beginning reconnection with the overlying flux at the top of the filament loop (TK05; HK16). Both cases match the terminal height and show the distortion. Case~1 shows it slightly more clearly and also reproduces the overall shape of the halted filament slightly better. The latter is quantified by the ratio of the leg crossing point height and the terminal height, which is 0.35 for \textsl{TRACE}, 0.27 for Case~1, and 0.24 for Case~2. 

Next, \textit{reconnected filament threads} appeared as the result of the first reconnection in the helical current sheet, which is formed by the helical kink instability in the interface between the flux rope and the ambient flux (Fig.~\ref{f:shape}(d)). The reconnection is strongest with the overlying flux and eventually cuts the whole flux rope. The figure shows this when about half of the flux in the rope is reconnected. A \textsl{TRACE} image that clearly shows the first two sets of reconnected threads (on the front and right-hand side) is used for the comparison. Both cases agree with the observations in the location of the main reconnection at the sides of the filament loop above the crossing point. This location is indicated by the bend points of the reconnected filament threads and flux rope field lines and by various intermittent brightenings in the full series of \textsl{TRACE} images (see the animations in \citealt{HJi&al2003}, \citealt{Alexander&al2006}, and HK16). The new footpoints of the two sets of reconnected filament threads lie in front of the other filament leg, relatively close to the middle of the whole structure. Both simulation runs show a similar location of the reconnected field line's new footpoints, with Case~1 approaching the observations somewhat better. 

The field line plots in Fig.~\ref{f:shape}(d) show a significant difference to the corresponding plots for the cases in HK16. Here, also the lower part of the erupted rope's legs reconnects with ambient flux. The resulting field lines are similar to those in the potential field of the TD99 magnetogram, different from the field lines that reconnect in the top part of the flux rope (see Figs.~6(c) and (e) in HK16). The additional reconnection here is due to the proximity of the ``sunspots'' (a smaller $L$), which strengthens the ambient flux low in the volume compared to the cases studied in TK05 and HK16. Only a small fraction of the flux in the rope reconnects in this way, so that the reformation of a flux rope (discussed next) is very similar to the previous cases. 

Finally, an arcade of \textit{flare loops} was formed (Fig.~\ref{f:shape}(e)). The simulations show this to result from a second phase of reconnection, which occurs in a vertical current sheet that forms between the legs of the cut flux rope near the crossing point in Fig.~\ref{f:shape}(c). This happens because the two halves of the cut rope are interlinked. This linking is a consequence of the writhing of the rising flux rope due to the helical kink (see Fig.~\ref{f:reconnection} below and HK16). Reconnection in a vertical current sheet is a key process in ejective events (eruptive flares/CMEs) and usually referred to as ``flare reconnection.'' It adds flux to the escaping flux rope above the current sheet and produces flare loops below the sheet. Here, the second reconnection reforms a flux rope \textit{below} the vertical current sheet and restores arch-shaped overlying flux
\textit{above} it, consistent with the unusual start height of the flare loops. The restored overlying flux shows an overall agreement with the observed flare loops in both cases. In particular, the ratio of projected height and footpoint distance, ${\approx}\,1.1$ for Case~1, and ${\approx}\,0.8$ for Case~2, matches the observed value of ${\approx}\,1.0$ quite well. Both cases also show that the restored overlying flux inherits a small part of the twist in the erupted flux rope, in agreement with the very unusual indication of twist in the flare loops. The field line plots additionally show sets of higher-arching loops that differ in shape from the flare loop arcade. These contain a minor part of the reformed overlying flux in Case~1, but a significant part (about one half) in Case~2. Hence, the comparison with the observed flare loops clearly favors Case~1 above Case~2. 

On the other hand, the value of $L$ in Case~2 lies closer to the distance of the flux concentrations from the polarity inversion line in the magnetogram of the eruption's source region. HK16 estimated the distance to be ${\sim}\,40$~Mm for both polarities. 

The sequence of the two major reconnection phases, the linking of the reconnected halves of the original flux rope, and the eventually resulting reformed flux rope are shown in more detail in Fig.~\ref{f:reconnection}. A flux rope is reformed in the location of the original one, which corresponds to the very low-lying bright EUV loops in Fig.~\ref{f:shape}(d). It contains nearly as much flux as the original rope, but is only weakly twisted, $\Phi<1\pi$.

\subsection{Rise profile and timing}

Figure~\ref{f:time_profile} shows the scaling of the two simulation runs to the observed rise profile of the eruption. Both cases can be scaled to agree quite well with the observational data. Case~1 fits slightly better, which can be best seen in the comparison of the velocities. 
The simulation time axes show that, in both runs, the instability needs a long time to lift the flux rope to the observed height, i.e., has a small growth rate. The instability in Case~1 (2) clearly begins to develop out of the initial noise ${\approx}\,10$ (${\approx}\,35$)~min before the initially very slow rise of the filament can be discerned in the\textsl{TRACE} data.  In fact both initial equilibria are chosen very close to the threshold of the helical kink mode. A small upward initial velocity perturbation, applied only at the flux rope apex and linearly rising to 0.04 (0.02)~$V_\mathrm{A0}$ at the termination time of 4 (2)~$\tau_\mathrm{A}$ for Case~1 (2), is required to launch the instability. The flux rope is slightly lifted above its initial position in the analytical (i.e., approximate) equilibrium during the perturbation and subsequent gradual development in the simulations. Thus, the scaled flux rope heights at 18:00~UT in Fig.~\ref{f:time_profile} lie slightly above the observed filament height. These heights match the observation much better than the ones in the cases analyzed in TK05 and HK16. More importantly, the more realistic inclusion of the line tying at the footpoints of the flux rope in the simulations analyzed here is not compromised by the slight initial lifting of the flux rope apex. 

The scaling determines the dimensional value of the Alfv\'en time. This combines with the dimensional value of the length unit $h_0$ to imply a value for the Alfv\'en velocity in the source region of the eruption, which is given in the plot and agrees with the range of known values of this parameter in active regions for both cases (see, e.g., \citealt{Innes&al2003}; \citealt{Wang&al2007}; \citealt{Regnier&al2008}).

\section{Discussion}

Both cases presented in this paper match the observations reasonably well, thus, both are a valid model of the event. Most items in the comparisons with the observational data yield only a small difference between the cases (mostly favoring Case~1); only the comparisons with the flare loop arcade and with the distance of the flux concentrations in the source region's magnetogram show moderate differences. We give the latter aspect a lower weight in our overall judgment, because the TD99 model in its original form employed here is generally not suited to closely model the complexity of the photospheric flux distribution in a solar active region. Overall then, Case~1 is the better-fitting run. Additionally, the smaller initial twist of this case is in better agreement with recent estimates of the twist that can accumulate in active regions before an eruption occurs \citep[see][]{YGuo&al2013,TLi&JZhang2015,Liu&al2016}.

Comparing Case~1 here with the best case of the broad parametric study of the event in HK16, we first note that both yield the plausible value of $\Phi=4\pi$ for the initial average twist. Both yield a good match of the observational data. None of the differences in the detailed matching is significant. Case~1 here is slightly superior in reproducing the distortion of the filament when the terminal height is reached and superior in matching the initial part of the rise profile, while the best case in HK16 is slightly superior in reproducing the inverse-gamma shape, the flare loop arcade, and the final part of the rise profile. The difference in matching the distance of the flux concentrations in the magnetogram (parameter $L=30$~Mm and 57~Mm vs.\ ${\sim}\,40$~Mm in the observational data) is not significant. Since Case~1 here achieves the agreement with the observations using a more realistic inclusion of the line tying, it represents the preferable model. 

A helical shape clearly develops in a number of solar eruptions. This is especially true for confined eruptions, but not restricted to them (e.g., \citealt{Romano&al2003}; \citealt{GZhou&al2006}; \citealt{RLiu&al2007}; \citealt{KSCho&al2009}; \citealt{Karlicky&Kliem2010}; \citealt{Joshi&Srivastava2011}; \citealt{BChen&al2014}; \citealt{ZXue&al2016}). Many of these events indicate the occurrence of the helical kink instability, especially those with a strong apex rotation \citep{Kliem&al2012}. The model pursued in the present paper should be applicable to them, and has indeed already shown good agreement for several cases. Further such events should be studied in the future. Alternative mechanisms for an apex rotation and the corresponding helical deformation of erupting flux, based on the action of an external shear field component \citep{Isenberg&Forbes2007} or on reconnection with ambient flux, tend to act on a much larger spatial scale of typically one solar radius or even more (\citealt{Lynch&al2009}; \citealt{Jacobs&al2009}; \citealt{Shiota&al2010}), whereas confined eruptions typically writhe strongly already in the low corona, as in the event studied here. On the other hand, a large number of confined eruptions does not show a strong writhing. Their explanation most likely requires a different mechanism.

\section{Conclusion}

The modeling of the confined filament eruption in solar active region NOAA 9957 on 2002 May~27 is here further improved beyond the results in TK05 and HK16 by including the photospheric line tying of the assumed initial flux rope in a more realistic manner. This is accomplished by using a flatter initial rope, whose apex height agrees exactly with the estimated filament height. A good matching of the observational data is found, equivalent to the previous modeling in the overall quantitative agreement, with only minor, non-significant differences in the details. 
Again, an initial average flux rope twist $\Phi\approx4\pi$ is indicated, and a less twisted flux rope is reformed by a sequence of two main reconnection phases, the first in a helical current sheet (different from ejective eruptions), the second in a vertical current sheet (partly similar to ejective eruptions). Additional reconnection of the flux rope legs with ambient flux reduces the flux content of the reformed flux rope somewhat in comparison to the simulation runs in HK16. 

The equivalent quantitative matching of the observational data by the improved model in this paper lends further support to models for solar eruptions, which assume an unstable flux rope to exist at the onset of the eruption's main acceleration phase. In particular, the model for strongly writhing confined eruptions by TK05 is substantiated, which assumes an initially kink-unstable flux rope in the stability domain of the torus instability.

\acknowledgements
We gratefully acknowledge helpful comments by R.~Santos de Lima and the referee. This work was supported by the German Academic Exchange Service (DAAD) and the German Science Foundation (DFG).

\bibliographystyle{an}
\bibliography{paper2}

\end{document}